# Anomalous behavior of membrane fluidity caused by copper-copper bond coupled phospholipids

Xiankai Jiang[1†], Jinjin Zhang[2†], Bo Zhou[3], Xiaojuan Hu[2], Zhi Zhu[1], Chao Chang[4], Junhong Lü[2*], Bo Song[1*]

[1]Terahertz Technology Innovation Research Institute, Shanghai Key Lab of Modern Optical System, Terahertz Science Cooperative Innovation Center, School of Optical-Electrical Computer Engineering, University of Shanghai for Science and Technology, Shanghai 200093, China

[2]Division of Physical Biology and CAS Key Laboratory of Interfacial Physics and Technology, Shanghai Institute of Applied Physics, Chinese Academy of Science, Shanghai 201800, China

[3]School of Electronic Engineering, Chengdu Technological University, Chengdu 611730, China

[4]Key Laboratory for Physical Electronics and Devices of the Ministry of Education, Xi'an Jiaotong University, Xi'an 710049, China

[†]These authors contributed equally to this work.

[*]Correspondence to: bsong@usst.edu.cn, lujunhong@sinap.ac.cn

**Abstract**

Membrane fluidity, well-known to be essential for cell functions, is obviously affected by copper. However, the underlying mechanism is still far from being understood, especially on the atomic level. Here, we unexpectedly observed that a decrease in phospholipid (PL) bilayer fluidity caused by $Cu^{2+}$ was much more significant than those induced by $Zn^{2+}$ and $Ca^{2+}$, while a comparable reduction occurred in the last two ions. This finding clearly disagrees with the placement in the periodic table of Cu just next to Zn and far from Ca. The physical nature was revealed to be a special attraction between $Cu^+$ cations, which can induce a motif forming of two phospholipids coupled by Cu-Cu bond (PL-*di*Cu-PL). Namely, upon $Cu^{2+}$ ion binding to a negatively charged phosphate group of lipid, $Cu^{2+}$ was reduced to $Cu^+$. The special attraction of the cations then caused one $Cu^+$ ion simultaneously binding to two lipids and another $Cu^+$, resulting in the formation of PL-*di*Cu-PL structure. In contrast, this attraction cannot occur in the cases of Zn and Ca ions due to their electron structure. Remarkably, besides lipids, the phosphate group widely exists in other biological molecules, including DNA, RNA, ADP and ATP, which would also induce the similar structure of Cu ions with the molecules. Our findings thus provide a new view for understanding the biological functions of copper and the mechanism underlying copper-related diseases.

**Significance Statement**

Membrane fluidity, critically essential for cell functions, is obviously affected by copper, but the underlying molecular mechanism is poorly understood. We propose that a special attraction between $Cu^+$ ions can cause a motif forming of two phospholipids (PLs) coupled by a Cu-Cu bond (PL-*di*Cu-PL), following $Cu^{2+}$ reduced to $Cu^+$ upon binding to PLs. The resulted motif of PL-*di*Cu-PL plays an unusual role in the fluidity of lipid bilayer. Our findings provide a new view for understanding biological functions of copper.

**Introduction**

Proper fluidity of the biological membrane is critically essential for numerous cell functions, such as adapting to the thermal stress of the environment of the microorganism (1), the binding of peripheral proteins associated at the lipid surface (2), reaction rates of enzymes (1), and even cell signaling and phagocytosis (3). Both *in-vivo* and *in-vitro* evidences have indicated that copper, as a biologically trace element, plays an important role in the membrane, especially in regard to its fluidity (4-8). However, the underlying mechanism is still far from being understood, partially because researches have been majorly devoted to the interactions of alkali and alkaline earth metal ions with phospholipids as well as the influences on the lipid bilayer (9-18). Traditionally, the effect of metal ions on the membrane was majorly attributed to the electrostatic attraction with lipid headgroups (19). This can explain the impact of divalent metal ions on membrane fluidity more than that of monovalent ones but cannot be applied to the differences of the influences of divalent ions, such as $Ca^{2+}$, $Mg^{2+}$, $Zn^{2+}$ and $Cu^{2+}$. Recently, Cremer and his coworkers studied effects of the $Cu^{2+}$ ion on a bilayer comprised of both phosphatidylcholine (PC) and phosphatidylserine (PS), and proposed that the ion was specifically bound to PS (20). Meanwhile, this binding was only stable under basic conditions, but not at acidic pH values. Further investigations suggested that a complex of $Cu(PS)_2$ formed upon $Cu^{2+}$ binding to PS molecules, which did not alter the net negative charge on the membrane (21). This differed from the manner and impact of $Ca^{2+}$ or $Mg^{2+}$ binding. Very recently, it was determined that the *cis* isomer of the $Cu(PS)_2$ complex was preferred to the *trans* one (22). Additionally, a synergetic effect of $Cu^{2+}$ with $Ca^{2+}$ were proposed, which potentially triggered the transition of PS membrane from fluid phase to soft solid phase (23). Besides those, influences of the $Cu^{2+}$ ion on a bilayer consisting of PC and phosphatidylethanolamine (PE) have also been explored (24). It was suggested that $Cu^{2+}$ could stably bind to the amine moieties of PE lipids, while other transition metal ions to PE bound in a similar manner. Noticeably, all these $Cu^{2+}$-lipid interactions specifically relate to the existence of amine moiety in the headgroup of the lipid.

Here, we propose an amine-independent copper-phospholipid motif in the membrane and apply it to illuminate our measurements of the anomalous effect of copper on the fluidity of a bilayer composed of PC and phosphatidylglycerol (PG). We observed that the decrease of the PC/PG bilayer fluidity caused by $Cu^{2+}$ ions was much more significant than those induced by $Zn^{2+}$ and $Ca^{2+}$ ions, while a comparable reduction occurred in the last two cases. A model of two phospholipids coupled by a Cu-Cu bond (*di*Cu) was built to explain the unexpected behavior of the bilayer fluidity induced by the $Cu^{2+}$ ion. Namely upon the interaction of two $Cu^{2+}$ ions with two phospholipids, one ion preferred simultaneously binding with the phosphate groups of the lipids and another ion after the $Cu^{2+}$ ions were reduced to $Cu^+$. The underlying physics was then revealed to be an anomalous $3d^{10}$-$3d^{10}$ attraction between $Cu^+$ cations, which was resulted from a special $3d^{10}$ closed shell of the outermost electron structure in $Cu^+$. In contrast, this attraction cannot occur in the cases of Zn and Ca ions due to their electron structures. Moreover, Ångström-resolution atomic force microscope (AFM) imaging also supported the formation of *di*Cu coupled to two lipids.

## Results and Discussions

**$Cu^{2+}$ ion caused anomalous fluidity of PC/PG bilayer.** The fluorescence recovery after photobleaching (FRAP) method involves the production of a concentration gradient of fluorescent molecules by irreversibly bleaching a portion of fluorophores in the observed region. The disappearance of this gradient over time is an indicator of the mobility of the fluorophores in the membrane as the fluorophore diffuses from the adjacent unbleached regions of the membrane into the bleached zone. We chose to use PC/PG/NBD (73:25:2) supported lipid bilayers for the FRAP measurements due to their large size, which allows their visualization in the microscopic field. The supported lipid bilayers appear large and uniformly fluorescent when observed through the confocal microscope. The inset in Fig. 1*A* shows fluorescence images of a representative FRAP experiment performed on such lipid bilayers: the dark circular region represents the bleached spot immediately after bleaching (0 s, bleach) and after the recovery of fluorescence at 80 s and 160 s (postbleach). The scanning parameters for all FRAP experiments were optimized to ensure no significant fluorescence photobleaching due to repeated imaging. Nonlinear curve fitting analysis of NBD fluorescence recovery kinetics after bleaching from the experimental data using the equations described in previous studies (25, 26) in the absence and presence of a 20 mM $CuCl_2$/$ZnCl_2$/$CaCl_2$ treatment on PC/PG bilayer mixtures are shown in Fig. 1*A*. The corresponding lateral diffusion rates of the NBD probe evaluated from Fig. 1*A* are displayed in Fig. 1*B*. The rate was 0.64 ± 0.03 for the sample incubated with $CuCl_2$, 1.12 ± 0.03 for $ZnCl_2$, 1.24 ± 0.03 for $CaCl_2$, and 1.80 ± 0.11 for the control (incubated with $NaCl_2$). These data indicate that the interaction of $Cu^{2+}$/$Ca^{2+}$/$Zn^{2+}$ ions with a PC/PG bilayer suppress the motion of the membrane. This can be attributed to the fact that divalent metal ions bind with the negatively charged lipid PG more stably than $Na^+$ (the control) (19), hence decreasing the electrostatic repulsion of the lipids, resulting in a close packing of lipid molecules. Subsequently, the constraints imposed on the displacement of lipids due to their enhanced order in the presence of metal ions in membranes lead to a reduced rate of lateral diffusion.

Remarkably, an unexpected order (Fig. 1*B*) of metal-ion impacts on the diffusion rate of lipids in the membranes was observed: $Cu^{2+}$ > $Zn^{2+}$ ~ $Ca^{2+}$ > $Na^+$ (control). This indicates that $Cu^{2+}$ plays a distinguished role in suppressing the mobility of lipids, more than the other ions, while a comparable effect occurs with the incubations of

Zn$^{2+}$ and Ca$^{2+}$. This result is obviously inconsistent with the placement in the periodic table of Cu just next to Zn and far from Ca.

**Analyses of mechanism underlying the anomalous fluidity of the bilayer.** To illustrate the mechanism under the anomalous influence of the Cu$^{2+}$ ion on the phospholipid membrane, we have studied the interactions of two Cu ions with two PG molecules. H$_3$C-[PO$_4$]$^-$-CH$_3$, with a phosphate group as a large portion of its composition, was employed as a simplified model of the phospholipid (PL). [Cu(H$_2$O)$_5$]$^{2+}$ was applied to simulate the Cu ion because it was found to coordinate five water molecules in solution (27).

First, two hydrated Cu$^{2+}$ ions interact with two phospholipids, respectively. We called the resultant state "State I" (Fig. 2C), denoted by [PL-Cu(aq)]$^+$, in which the hydrated Cu cation bound the negatively charged oxygen atom in the phosphate group of the lipid. The label aq stands for the water molecules in the hydrated group. Using Eq. 1, the binding energy of a Cu$^{2+}$ ion in this state was calculated by an *ab initio* method based on density functional theory (DFT) with the solvation effect of the outer water environment,

$$\begin{aligned} E_{binding}^{I}(\text{Cu}) &= \frac{1}{2}\left[2E([\text{PL-Cu(aq)}]^+) - 2E(\text{PL}^-) - 2E(\text{Cu(aq)}^{2+})\right] \\ &= E([\text{PL-Cu(aq)}]^+) - E(\text{PL}^-) - E(\text{Cu(aq)}^{2+}). \end{aligned} \quad (1)$$

$E(\text{PL}^-)$, $E(\text{Cu(aq)}^{2+})$ and $E([\text{PL-Cu(aq)}]^+)$ indicate energies of the phospholipid, hydrated Cu$^{2+}$ ion and their binding state I, respectively. As presented in Fig. 3A, the binding strength reached -37.14 kcal/mol, meaning that the hydrated Cu ion can bind to the oxygen of the phosphate group. Natural-bond-orbital (NBO) analysis (28) showed that the Wiberg bond order (29) was 0.374 for the Cu-O bond (Table S1 in *SI*), suggesting a chemical bond is occurring with a few covalent characteristics. This chemical bond can consequently provide the stable binding of the Cu ion with the O of phospholipid in solution.

Second, two hydrated $Cu^{2+}$ ions simultaneously bind with two phospholipids, resulting in a *di*Cu coupled lipid pair (Fig. 2D). The resulting conformation, referred to "State II", had two positive charges (denoted by $[PL\text{-}diCu(aq)\text{-}PL]^{2+}$). The optimized Cu-O and Cu-Cu bond lengths in the resulting PL pair were 1.90 Å and 2.58 Å, respectively. The frequencies of Cu-related vibration modes were $0.4 \times 10^{10}$ Hz, $1.7 \times 10^{10}$ Hz and $3.1 \times 10^{10}$ Hz. The binding energy of a $Cu^{2+}$ ion in State II was calculated via Eq. 2 as follows,

$$\begin{aligned}E_{\text{binding}}^{\text{II}}(\text{Cu}) &= \frac{1}{2}\Big[E([PL\text{-}diCu(aq)\text{-}PL]^{2+}) - 2E(PL^-) - 2E(Cu(aq)^{2+})\Big] \\ &= E([PL\text{-}diCu(aq)\text{-}PL]^{2+})/2 - E(PL^-) - E(Cu(aq)^{2+}),\end{aligned} \quad (2)$$

where $E([PL\text{-}diCu(aq)\text{-}PL]^{2+})$ denotes the energy of the copper-phospholipid complex in State II. Surprisingly, the binding strength of a Cu ion in PL-*di*Cu-PL reached -38.88 kcal/mol, which is surprisingly greater than the strength of -37.14 kcal/mol in State I. Moreover, the Wiberg bond order was 0.421 for the Cu-Cu binding in State II (Fig. 2B), suggesting that a chemical bond with a definite covalent characteristic is formed (more information shown in the following part with Fig. 4). All of these results suggest that the Cu ions can also bind to the phospholipids in State II, similar to in State I, and even more stably than in State I with the presence of water.

It should be noted that State II [PL-*di*Cu-PL]$^{2+}$ can be taken as the coupling of two groups [PL-Cu]$^{+}$ (the structure in State I) (see Fig. 2*C,D*). The positive charges of the two groups would cause a Coulomb repulsion and subsequently hinder the formation of State II. Hence, we performed an electron-structure analysis to reveal the physics underlying the formation of State II. The Cu ion in the structure [PL-*di*Cu-PL]$^{2+}$ had an NBO charge of only +0.97 e, indicating that the Cu$^{2+}$ ion is reduced to Cu$^{+}$ with an electron configuration of [Ar]3d$^{10}$4s$^{0}$ upon binding to the negatively charged phospholipid. Additionally, electrons were observed at the midpoint of two Cu ions with a density of 0.043 e/Å$^{3}$ (Fig. 4*A* left, and Table S2 in *SI*). The pattern of electron localized function (ELF) was approximately square in the region close to the Cu ion (the yellow squares under the yellow balls in the lower-left region of Fig. 4*B*), suggesting that the outermost electrons majorly occupy the 3d orbital. The ELF value was 0.254 at the midpoint of two Cu ions (Fig. 4*B* upper-left), indicating that there exists an electron pair localized in this area with a probability of 25.4%. Further NBO analysis showed a chemical bond clearly occurring between the two Cu ions (Fig. 4*C* left). The bond was composed of 50% of the valance orbitals of each Cu ion, in which the ratio of Cu 3d orbital reached 84.3%. Thus, the 3d electrons of the Cu ions substantially contributed to the Cu-Cu bond. All of these results suggest that a strong 3d$^{10}$-3d$^{10}$ attraction (30, 31) occurs between the ions after the Cu$^{2+}$ is reduced, upon binding to the negatively charged phosphate group of lipid. Therefore, it is this attractive force that suppresses the Coulomb repulsion between two [PL-Cu]$^{+}$, resulting in the formation of [PL-*di*Cu-PL]$^{2+}$. Notably, the 3d$^{10}$-3d$^{10}$ attraction of the $Cu^+$ ions has already been reported and applied in inorganic materials (32-34).

For comparison, we have also studied the binding characteristics of other metal ions M (M = $Na^+$, $K^+$, $Mg^{2+}$, $Ca^{2+}$, $Zn^{2+}$ and $Ag^{2+}$) with $H_3C$-$[PO_4]^-$-$CH_3$ fragments. It was found that only $Ag^{2+}$ presented the same behavior as $Cu^{2+}$ upon binding to the phospholipids, while the others did not show this manner. We first calculated the binding energies of hydrated M ions with $H_3C$-$[PO_4]^-$-$CH_3$ in States I and II, respectively. The results are presented in Fig. 2A. For State I, the binding energies were -23.14 kcal/mol for $Na^+$(aq), -22.37 kcal/mole for $K^+$(aq), -33.86 kcal/mol for $Mg^{2+}$(aq), -26.14 kcal/mol for $Ca^{2+}$(aq), -33.29 kcal/mol for $Zn^{2+}$(aq), and -38.94 kcal/mol for $Ag^{2+}$(aq). These results indicate that all of these ions can bind to the phosphate group of a lipid in State I, and the binding of divalent ions is more stable than that of monovalent ions. For State II, negative values were observed only in the binding energies of the hydrated $Ag^{2+}$ ion (-45.10 kcal/mol) and $Zn^{2+}$ ion (-22.13 kcal/mol), and not in the other ions. Moreover, a larger binding strength in State II than in State I occurred with the $Ag^{2+}$ ion but not with the $Zn^{2+}$ ion. Additionally, the Wiberg bond order of M-M binding was 0.337 for the Ag ions in State II and was less than 0.1 for the others, suggesting that a chemical bond only occurs in a PL-*di*Ag-PL structure with a covalent characteristic, but not in the cases of other metal ions. Therefore, with the competition of State I, State II is stable only for $Ag^{2+}$ ions in the presence of water, but not for the other ions. The NBO charge of the $Ag^{2+}$ ion was +0.93 in State II, denoting that $Ag^{2+}$ is reduced to $Ag^+$ upon binding to the phosphate group of lipid, and then, a closed shell [Kr]$4d^{10}5s^0$ of electron configuration occurs in the Ag ion. Therefore, the formation of the $Ag^+$-$Ag^+$ chemical bond can be attributed to the $4d^{10}$-$4d^{10}$ attraction, similar to the case of [PL-*di*Cu-PL]$^{2+}$. We thus conclude that the similar behavior of Ag ion and Cu ion in State II results from the structure of valence electrons, which is the same for Group 11 metals. This special property consequently leads to the distinguished difference of these ions with $Zn^{2+}$, $Ca^{2+}$, $Mg^{2+}$, $K^+$ and $Na^+$ ions upon binding to phospholipids.

Based on the previous DFT calculations, we have improved the CHARMM36 force field specifically for Cu-Cu and Cu-O bindings in PL-*di*Cu-PL structure (35) (detailed information presented in Table S3 and the corresponding discussion of *SI*) and performed classical molecular dynamics simulations (MD) to explore the influences of the Cu-containing structure on lipid bilayer. A molecular model, consisting of two PG molecules coupled by two $Cu^+$ ions (Fig. 5*A*), was employed in the MD. The positive charge of the PL-*di*Cu-PL structure was neutralized by $Cl^-$ ions in the presence of water molecules. Through a 1.0-μs simulation of each sample, the lipids assembled to form a pattern of stripes (Fig. 5*B-C*). A segment of assembled PL-*di*Cu-PL structures are shown in Fig. 5*D*. Two $Cl^-$ ions were observed over and under the Cu-O plane, respectively, indicating that $Cl^-$ ions are involved in the assembly of the membrane due to an electrostatic attraction. To study the order of the degree of assembled lipids in a membrane, we applied three parameters: the angles of O-O direction in phospholipid along the x ($\theta$) and z ($\varphi$) axes (upper-left inset of Fig. 5*E*) for the orders parallel to the membrane surface and the position ($d_z$) of the phosphorus atom in the membrane along the z-axis for the order perpendicular to the surface. The coordinate origin of the z-axis was set at the midpoint of two layers. First, in the distribution of lipids according to the parameters $\theta$ and $\varphi$ with presence of Cu ions (the orange curves in Fig. 5E), two peaks were observed at $\theta = \pm 60.5°$, and $\varphi = \pm 42.5°$. In the distribution according to $d_z$, two peaks were located at approximately $\pm 2.0$ nm. The plus/minus symbols above resulted from two opposite O-O directions in the PL-$di$Cu-PL structure (Fig. 5A) for the cases of the angles $\theta$ and $\varphi$ and from the two layers of the membrane for $d_z$. In contrast, without the presence of Cu ions, there was no peak for $\theta$, and a broad peak for $\varphi$ (at -90° or +90°, due to a periodic condition), clearly suggesting that the lipids are out of order in the direction parallel to the membrane surface. The full-width half-maximum (FWHM) of the peak was then employed to quantitatively study the order. The FWHM of the $\theta$-related peak was 180° (no peak) without Cu ions and 42° with Cu ions. The reduction from 180° to 42° obviously indicates that Cu ions induce a change in lipids from a disordered state to an ordered state. Similar observations occurred for the other parameters. The FWHM of $\varphi$-related peak decreased from 63° (without Cu ions) to 25° (with Cu ions), and the FWHM of $d_z$-related peak was reduced from 0.7 nm (without Cu ions) to 0.4 nm (with Cu ions). Moreover, the narrow peaks (FWHM = 25°) at $\varphi = \pm 42.5°$ with Cu ions further suggested that for the assembled lipids, most of the Cu-O planes were parallel to each other, with an angle of approximately 42.5° (Fig. 5D). We thus conclude that Cu ions can clearly enhance the order of lipids in a membrane, especially in the directions parallel to the membrane surface. Roughness of the membrane surface has also been calculated by the mean square deviation (MSD) of phosphorus-atom displacements along the z-axis of the membrane. The values in the presence and absence of $Cu^{2+}$ ions were $0.20 \pm 0.02$ nm and $0.26 \pm 0.04$ nm, respectively (Fig. 5$F$). The significant difference between them obviously suggests that Cu ions suppress the roughness of membrane. Moreover, AFM measurements further supported the theoretical results above. As shown in Fig. 5$G$-$H$, a pattern of clear stripes was observed in the $CuCl_2$-incubated PC/PG bilayer and not in the control (incubated with $NaCl_2$). The roughness of the membrane was $0.12 \pm 0.01$ nm for the presence of $CuCl_2$ and $0.11 \pm 0.08$ nm in its absence, and the P value between them was 0.015. The AFM data above indicate that Cu ions can significantly suppress the roughness of a lipid bilayer. It is noted that the difference of roughness values between AFM and MD results can be attributed to that a substrate for lipid bilayer was applied in AFM measurements but not in MD simulations.

From both the theoretical and experimental analyses, we conclude that a pattern of stripes can be assembled in the membrane with help of the special Cu-lipid interaction and the Cl ions involved. This significantly enhanced order then imposes a strong constraint on displacement of lipid molecules in the bilayer, clearly reducing the fluidity of membrane, which results in the anomalous effect of $CuCl_2$ previously observed by FRAP measurements.

**Conclusion**

In summary, we proposed a motif of *di*Cu coupled phospholipids, which can illuminate the anomalous decrease of membrane fluidity caused by $Cu^{2+}$, compared to those by $Zn^{2+}$ and $Ca^{2+}$. The mechanism under the motif formation was further revealed. Upon the $Cu^{2+}$ ion interacting with the lipid, $Cu^{2+}$ was reduced to $Cu^{+}$. After that, one $Cu^{+}$ ion preferred simultaneous binding to two phospholipids and another $Cu^{+}$, due to the anomalous $3d^{10}$-$3d^{10}$ attraction between the metal ions. In contrast, this attraction cannot occur in the cases of Zn and Ca ions due to their electronic structures. It is worth noting that besides lipids, the phosphate group also widely exists in other biological molecules, such as DNA, RNA, ADP, ATP, and enzymes. Therefore, as a kernel of the motif, the structure of *di*Cu coupled phosphate groups and the anomalous Cu-Cu attraction (see Fig. 2*D*) will provide a new direction for understanding the biological function of copper as a trace element essential to our life, as well as to the mechanisms of copper-related clinical diseases.

## Materials and Methods

**Preparation of lipid bilayers for confocal microscope imaging.** The supported lipid bilayers were prepared from the negatively charged lipid DOPG (1,2-dioleoyl-sn-glycero-3-(phospho-rac-(1-glycerol))), neutral lipid DOPC (1,2-dioleoyl-sn-glycero-3-phosphocholine) and headgroup labeled NBD-PE (1,2-diphytanoyl-sn-glycero-3-phosphoethanolamine-N-(7-nitro-2-1,3-benzoxadiazol-4-yl)). DOPC, DOPG and NBD-PE solutions in chloroform were mixed to achieve a DOPC/DOPG/NBD-PE molar ratio of 73:25:2; the solvent was evaporated under nitrogen and the dried lipid film was suspended in TBS buffer (50 mM Tris, 150 mM

NaCl) to a concentration of 1 mM. The lipid suspension was then sonicated to clarity, yielding a suspension of small unilamellar vesicle liposomes. The small unilamellar vesicle suspension was then exposed to a clean glass surface (a microscope petri dish was first etched by plasma (Harrick, PDC-32G, 4 minutes in air/vacuum) then was cleaned by 1% hydrofluoric acid and thoroughly rinsed with deionized water and dried under nitrogen) and incubated for 1 hour at room temperature to form lipid bilayers. The excessive unfused liposomes were removed from rinsing with excess of the buffer. The 20 mM $CuCl_2/CaCl_2/ZnCl_2$ solution was added to the lipid bilayers before imaging.

**Confocal microscope experimental setup and data acquisition.** A commercial confocal microscope (Leica TCS SP5) was used for the fluorescence recovery after photobleaching (FPAR) measurements. A 488-nm Argon laser was used as the excitation source. The sample was illuminated and the fluorescence emission was collected by a 63× oil immersion objective. The dimensions of the acquired regions were 180 μm × 180 μm. The bleaching pulse was applied by rapidly scanning a focused laser beam over an area with a dimension of 8 μm × 8 μm for 30 s with an interval of 1s at full laser power. Immediately after bleaching, the region of 180 μm × 180 μm was recorded for 300 s with an interval of 5 s at low excitation energy.

*Ab initio* **calculations.** Our *ab initio* calculations based on the density functional theory (DFT) as well as the electron structure analyses were implemented in the Gaussian09 package (36). The geometry optimizations and vibrational frequencies of all compounds were carried out at DFT level, employing the M06L functional (37). A mixed basis set GEN (SDD basis sets for Cu, Zn and Ag atoms, and 6-31+G(d,p) set for other atoms) was applied for all the calculations in this study. The optimized stationary points were identified as minima or first-order saddle points. Solvation effects of outer water environment were taken into account by calculating the single-point energies of the optimized configurations under the integral-equation-formalism polarizable continuum model (IEFPCM) of solvation (38) at the same level of theory as used in the gas-phase optimizations. To investigate the coordination effects on bond strength and charge distribution, the natural-bond-orbital (NBO) method was used for all complexes. Orbital populations and Wiberg bond orders were calculated with the NBO 3.0 program implemented in Gaussian 09.

**MD simulations.** The initial configuration of lipid bilayer system was generated by MemBuilder server (39). A total of 128 DMPG lipid molecules were placed periodically in each lipid bilayer, and the number of water molecules per lipid was 45. According to the optimized structure obtained from the DFT calculations, the headgroup of two adjacent lipid molecules were linked by the Cu-O (0.1906 nm) and Cu-Cu (0.2584 nm) bonds with the angle O-Cu-O (172.53°). The harmonic potential force constant for the bond Cu-O, Cu-Cu stretching and the bond-angle O-Cu-O vibration are 483660 kJ mol$^{-1}$nm$^{-2}$, 135450 kJ mol$^{-1}$nm$^{-2}$ and 800 kJ mol$^{-1}$rad$^{-2}$, respectively. A corresponding number of Cl$^-$ ions was added to neutralize the system.

We performed the MD simulations for the system relaxation in an NPT ensemble at high temperature 320 K for 600 ns. After that, we selected one conformation per 20 ns in the time interval from 500 ns to 600 ns, and obtained five samples as initial structures. Finally, The MD simulations were performed in an NPT ensemble at the temperature of 303 K with 1.0-μs for each sample.

All simulations were performed using GROMACS 5.1 (40) with a time step of 2 fs. The CHARMM36 force field for lipids (41, 42) and the CHARMM TIP3P water model (43) were used. The particle mesh Ewald (PME) method (44, 45) was used to treat long-range electrostatic interactions, whereas the van der Waals interactions were treated with a 1.0 nm - 1.2 nm force-based switching function (46). The temperature was maintained at 303 K using Nosé-Hoover thermostat (47, 48) with a coupling constant of 1 ps, and the pressure was kept constant at 1 bar using semi-isotropic Parrinello-Rahman barostat (49) with a coupling constant of 5 ps and a compressibility of $4.5 \times 10^{-5}$ bar$^{-1}$. Periodic boundary conditions were applied in the three directions. After a series of minimization and equilibration steps suggested by CHARMM-GUI (46), the data were collected every 2 ps during the next 1 μs production run.

**Preparation of Lipid Bilayers for AFM imaging.** The negatively charged lipid, 1,2-Dimyristoyl-sn-glycero-3-phosphorylglycerol (DMPG), and neutral lipid, 1,2-Dimyristoyl-sn-glycero-3-phosphorylcholine (DMPC) were purchased from Avanti Polar Lipids (Alabaster, AL) and used without further purication. Lipid bilayers on freshly cleaved mica surface were prepared following the vesicle fusion method. Briefly, DMPC/DMPG (3:1) mixtures were first dissolved in chloroform, followed by evaporation of the solvent under nitrogen. After that lipid mixtures were dissolved in 50 mM Tris, 150 mM NaCl to a concentration of 1.5 mg/ml and sonicated in a bath sonicator until clear to form small vesicles. A 20 μL droplet of the vesicle solution was then applied to a freshly cleaved fragment of mica, incubated for about 2 h at room temperature, and then the sample was incubated at 35 °C for 40 min to fluidize the lipid, a necessary step to form the bilayer.

**AFM Imaging.** The sample was placed within the AFM (Nano III, Veeco) and imaged in the contact mode using DNP tips (Bruker), with a spring constant of 0.06 N/m. The scan rate was 10 Hz and the applied force was minimized to about 0.1 nN.

**Acknowledgments.**

This work is supported by National Natural Science Foundation of China (Nos. 11474298, 11405250), Key Research Program of Frontier Sciences of the Chinese Academy Sciences (No. QYZDJ-SSW-SLH019) and the Tianjin Supercomputer Center of China.

**Author Contributions**

B.S. and J.L. conceived the idea and jointly supervised the project. J.L. designed the experiments. J.Z. and X.H. carried out the experiments, J.L. and B.S. performed the data analysis. The theoretical simulations and calculations were designed by B.S., carried out by X.J., B.Z, Z.Z. and B.S.. B.S. performed the theoretical analysis. C.C. performed some analysis. B.S. and J.L. co-wrote the paper. All authors discussed the results and commented on the manuscript.

**Competing interests:** The authors declare no competing financial interests.

**Figure Legends**

**Fig. 1.** Fluidity of a lipid bilayer in the presence of metal ions. (*A*) Normalized fluorescence intensity. The red circle, green triangle, blue inverted triangle and black rectangle indicate the fluorescence data with incubation of $Cu^{2+}$, $Zn^{2+}$, $Ca^{2+}$ and control ($Na^+$), respectively. The data are fitted with the curves with corresponding colors. The insets are the fluorescence-recovery images with incubation of $Cu^{2+}$ at the times 0 s, 80 s and 240 s. (*B*) Rates of fluorescence recovery. The rate of incubation with $CuCl_2$ is obviously less than those with other buffers, while the rates with $CaCl_2$ and $ZnCl_2$ are comparable (dark-red dashed line).

**Fig. 2.** Ways of two hydrated $Cu^{2+}$ ions binding with two phospholipids. The cyan, brown, red, white and yellow balls represent carbon, phosphorus, oxygen, hydrogen and copper, respectively. (*A*) A phosphate group in a phospholipid (PL). (*B-D*) A simplified model, $H_3C$-$[PO_4]^-$-$CH_3$, of a phospholipid for studying the interaction with $Cu^{2+}$ ions. (*B*) Initial state. Four molecular groups (two negatively charged fragments $H_3C$-$[PO_4]^-$-$CH_3$ and two hydrated copper cations $[Cu(H_2O)_5]^{2+}$) are separated by a large distance. (*C*) State I. Two hydrated $Cu^{2+}$ ions bind to two phospholipids, respectively, which results in two $[PL-Cu(aq)]^+$ structures. (*D*) State II. Two hydrated $Cu^{2+}$ ions bind simultaneously with two phospholipids, forming a $[PL\text{-}diCu(aq)\text{-}PL]^{2+}$

structure.

**Fig. 3.** Binding energies (*A*) and bond orders (*B*) for hydrated metal ion $M^{n+}$ in States I and II. $n = 1$ for M = Na and K, while $n = 2$ for M = Mg, Ca, Zn, Cu and Ag. The bond orders (*B*) suggest that a chemical bond with a covalent characteristic occurs for Cu and Ag and an ionic bond occurs for the other metals. Moreover, the binding behavior of $Cu^{2+}$ in State II significantly differs from that of $Zn^{2+}$, although Zn is the element just next to Cu in the periodic table.

**Fig. 4.** Electron analyses of Cu ions (left column) in State II by comparison with Zn ions (right column). The cyan, brown, red, white, yellow and dark gray balls represent carbon, phosphorus, oxygen, hydrogen, copper and zinc, respectively. (*A*) Electron densities. The green cloud denotes the electron density with an isosurface of 0.04 e/Å$^3$. (*B*) Electron localized function (ELF). Upper: One-dimensional ELF along the metal-metal direction. The gray area indicates the region of core electrons, where ELF decays and quickly vanishes because a pseudo potential is employed in DFT calculations. Lower: Two-dimensional ELF. The pattern of ELF is approximately square in the area close to a Cu ion (lower-left) and is a circle in the area close to a Zn ion (lower-right). These suggest that the outermost electrons majorly occupy the 3d orbital for Cu and the 4s orbital for Zn. (*C*) Natural bond orbital between metal ions. The orange and light blue clouds indicate the orbital with an isosurface of 0.04 e/Å$^3$.

A metal-metal bond orbital occurs for Cu ions but not for Zn ions. For clarity, water molecules are not shown.

**Fig. 5.** Assembly and order-analyses of phospholipids in membranes caused by Cu ions. (*A-F*) Lipid assembly induced by PL-*di*Cu-PL structure through MD simulations. The light blue, brown, red, yellow and silver balls represent carbon, phosphorus, oxygen, copper and chlorine, respectively. For clarity, hydrogen atoms and water molecules are not shown. (*A*) Applied PL-*di*Cu-PL structure in the PG-bilayer simulations. (*B-C*) Typical conformation of assembled lipids in membrane with side (*B*) and top (*C*) views. The blue box represents the periodic boundaries applied in the simulations. The tails of lipids and the atoms in the terminal of the lipid heads are represented by light blue curves and blue balls, respectively, for clarity. (*D*) A segment of assembled lipids. Two Cl$^-$ ions (silver balls) are observed over and under (dotted lines) the Cu-O plane, respectively. (*E-F*) Analyses of PL-*di*Cu-PL induced assembly based on the degrees of order (*E*) and roughness (*F*). (*E*) The upper-left inset shows the applied angles ($\theta$, $\varphi$) for the order degree in x-y plane. Upper and middle: Distributions of lipids according to the angles $\theta$ and $\varphi$, respectively. The black dotted lines denote the locations ±42.5° of two peaks. Lower: Distribution of phosphorus atoms in lipids along the z direction ($d_z$), where the coordinate origin is set at the midpoint of two layers. The full-width half-maximum (FWHM) is presented for peaks in presence (orange value) and absence (blue value) of Cu ions. (*G-I*) AFM measurements. AFM images without (*G*) and with (*F*) incubation of CuCl$_2$, and a comparison of the roughness of membrane surfaces (*I*). (*H*) The light green curve

represents a stripe clearly in the membrane. (*I*) The label * over double-arrow line indicates the P value < 0.05 of significant difference between the two groups of data. This suggests a significant difference in roughness of the membranes with and without $CuCl_2$.

**Figures**

**Figure 1**

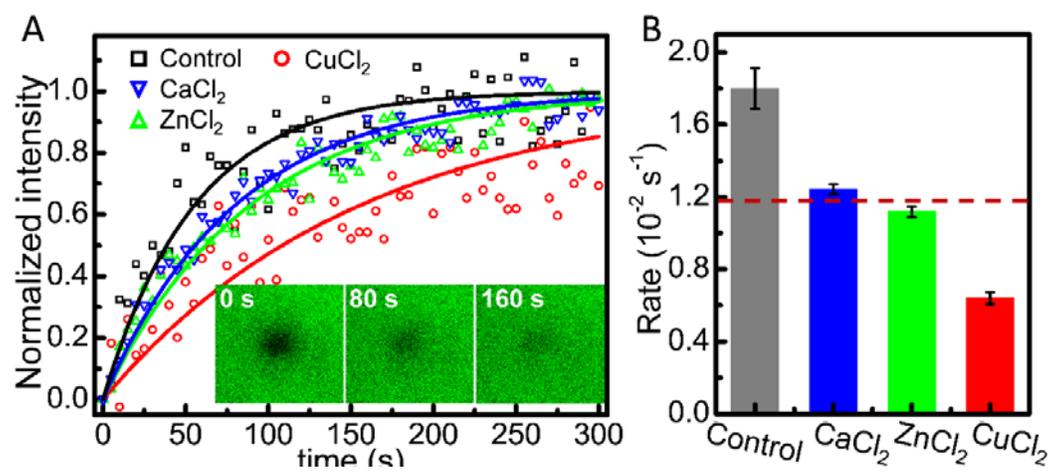

**Figure 2**

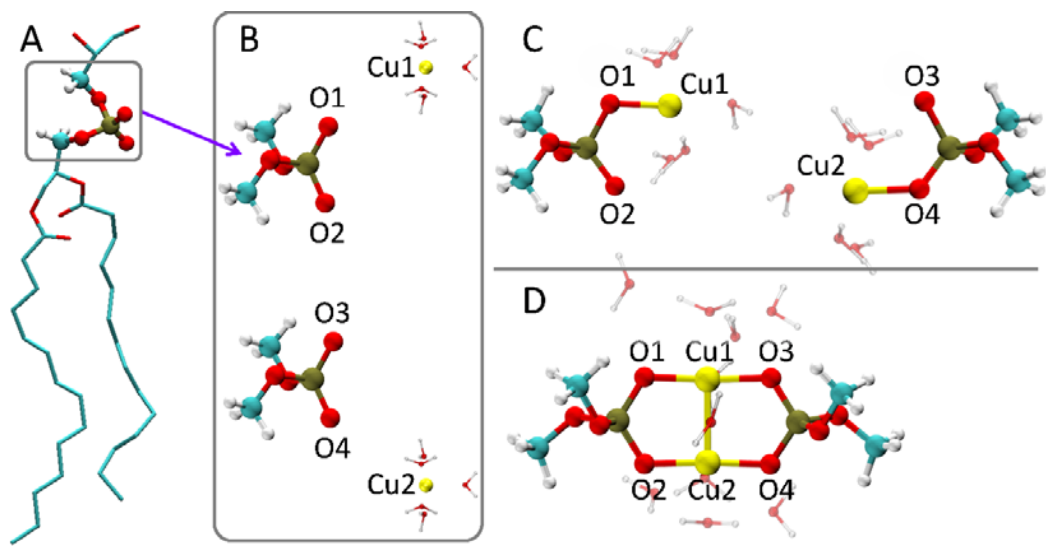

**Figure 3**

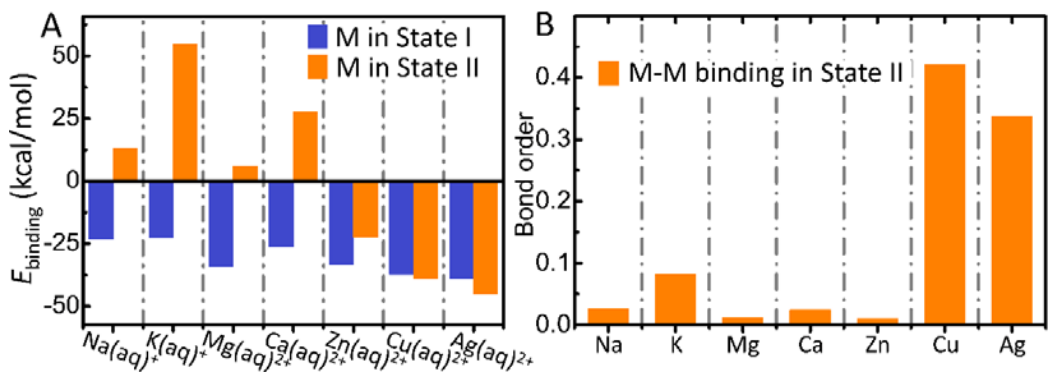

**Figure 4**

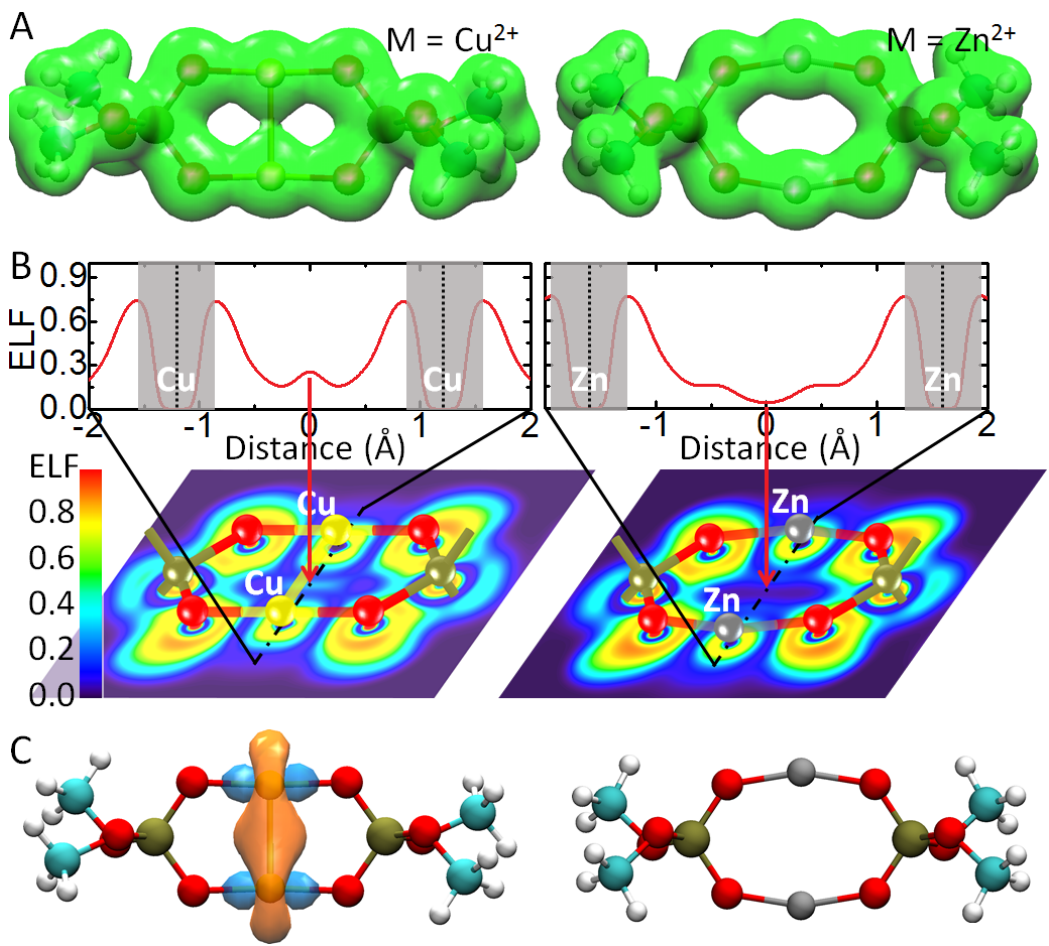

**Figure 5**

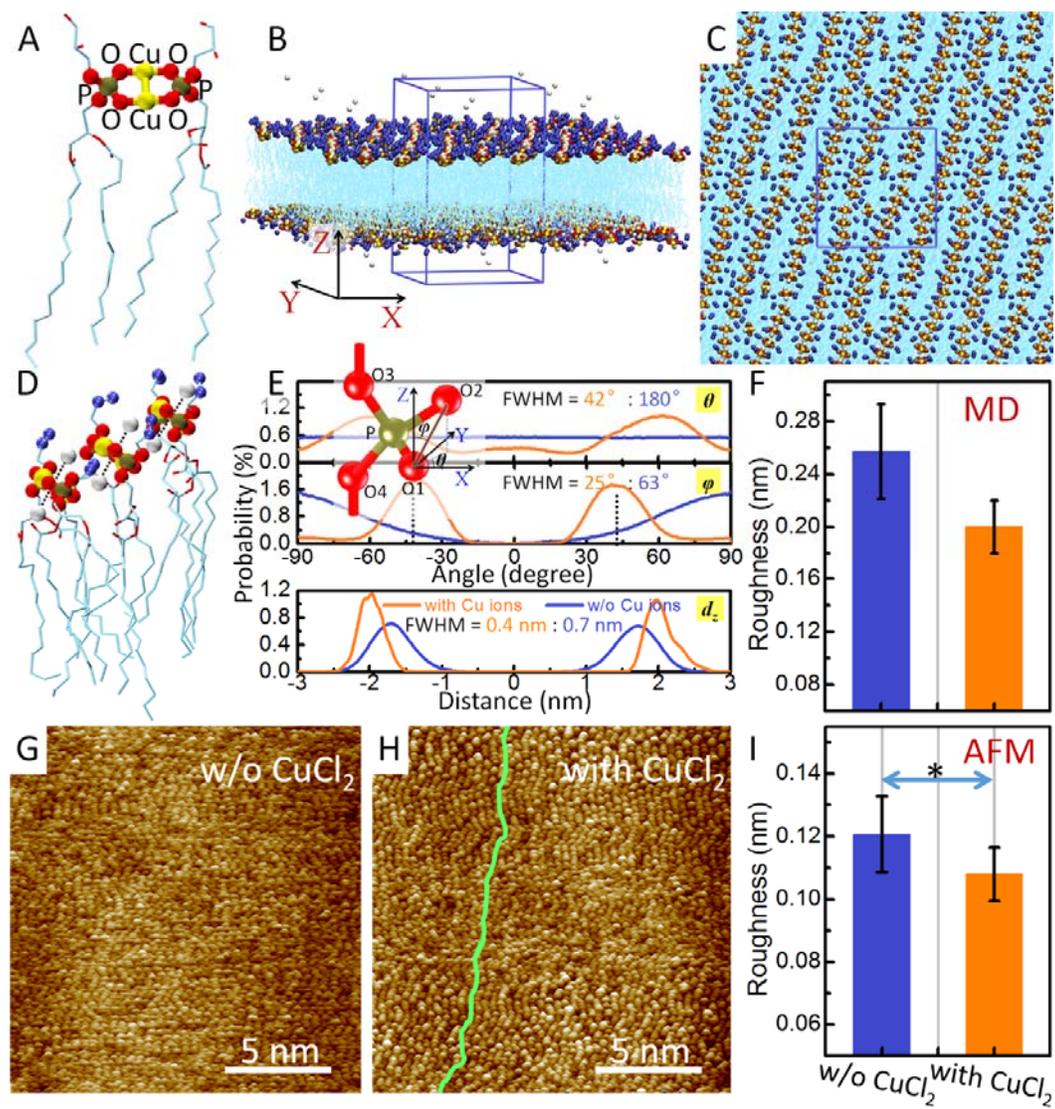